\documentclass{pasj00}
\usepackage[dvips]{color}

\begin{document}
\SetRunningHead{K. Otsuji et al.}{Statistical Study on Solar Flux
Emergence}
\Received{2010/12/28}
\Accepted{2011/05/17}
\Published{2011/10/25}

\title{Statistical Study on the Nature of Solar Flux Emergence}

\author{%
  Kenichi~\textsc{Otsuji},
  Reizaburo~\textsc{Kitai},
  Kiyoshi~\textsc{Ichimoto},
  and Kazunari~\textsc{Shibata}
  }
\affil{Kwasan and Hida Observatories, Kyoto University, yamashina-ku, Kyoto
607-8741}
\email{otsuji@kwasan.kyoto-u.ac.jp}


\maketitle

\begin{abstract}
We studied 101 flux emergence events ranging from small ephemeral regions to
large emerging flux regions which were observed with \textit{Hinode} Solar
Optical Telescope filtergram.
We investigated how the total magnetic flux of the emergence event controls
the nature of emergence.
To determine the modes of emergences, horizontal velocity fields of global
motion of the magnetic patches in the flux emerging sites were measured by the
local correlation tracking.
Between two main polarities of the large emerging flux regions with more than
around $2\times10^{19}$ Mx, there were the converging flows of anti-polarity
magnetic patches.
On the other hand, small ephemeral regions showed no converging flow
but simple diverging pattern.
When we looked into the detailed features in the emerging sites,
irrespective of the total flux and the spatial size, all the emergence
events were observed to consist of single or multiple elementary emergence unit(s).
The typical size of unitary emergence is 4 Mm and consistent with the
simulation results.

From the statistical study of the flux emergence events, the maximum spatial
distance between two main polarities, the magnetic flux growth rate and the mean
separation speed were found to follow the power-law functions of the total 
magnetic flux with the indices of 0.27, 0.57, and -0.16, respectively.
From the discussion on the observed power-law relations, we got a physical
view of solar flux emergence that emerging magnetic fields float and evolve
balancing to the surrounding turbulent atmosphere.
\end{abstract}

\section{Introduction}
The sites where the subsurface magnetic flux tubes emerge on solar surface are
called emerging flux regions \citep{Bru69, Zir72}.
The typical simple emerging flux region has one main pair of the opposite
magnetic concentrations at the both ends of the emerging site.
The opposite magnetic concentrations of the main pair move away from each other
with the speed of 1-2 km$^{-1}$ \citep{Zwa85, Bra85a} in the developing phase of the region.
Inside the developing emerging site there are many magnetic flux tube emerging
on the photosphere.
The newly emerged magnetic flux tubes at the emerging site appear as the dark
granular lanes \citep{Lou61, Bra85b} on the photosphere.
At the both ends of the dark granular lane, there are the magnetic
concentrations which are called the footpoints and observed as the bright
points in G band image \citep{Ots07}.
With H$\alpha$ the emerged flux tubes are observed as dark arch filaments
\citep{Bru67}.
The lifetime of an arch filament is 10-30 minutes \citep{Bru67, Cho88}.
The rise velocity of arch filaments is 10-15 km s$^{-1}$ \citep{Bru69,
Cho88}. \citet{Ots10} found the deceleration of the apex of the
small-scale arch filaments in the chromosphere.

Emerging flux regions show a variety of size, lifetime, total magnetic flux and
field strength.
Especially the small emerging flux regions are called ephemeral active region
(EAR; \cite{Har73}). For convenience, in the following page we define larger
(i.e. non-EAR) emerging flux regions as EFRs.
EFRs are produced by fairly large-scale flux emergence.
They have a pair or more complex group of sunspots with definite penumbrae.
The typical size of EFRs is more than 30 Mm \citep{Bru67}.
EFRs show their emergence activities for several days and exist on the solar
surface for a few month at the maximum.
The total flux in an EFR increases with the rate of $10^{20}$ Mx hr$^{-1}$
and reaches to the order of $10^{20}$-$10^{22}$ Mx \citep{Zwa87}.
In the fairly developed EFR the field strength of main spot is around 3,000
Gauss \citep{Bra82}.

EARs have simple bipolar configuration.
They have no penumbra in the sunspots.
The typical size varies from 5 Mm to 30 Mm \citep{Har73, Har75,
Hag01, Ots07}. EARs have short life time of hours or one day.
The total flux in an EAR is up to $10^{20}$ Mx with the increase rate of
$10^{19}$ Mx hr$^{-1}$.
The magnetic field strength of main spot is from a few times 100 Gauss
\citep{Mar88} to 2,000 Gauss \citep{Bra82}.

Various simulation studies on flux emergence have been performed by many
researchers \citep{Shi89, Mat92, Fan01}.
They showed the simple bipolar emergence simulation which corresponds to the
observation result.
\citet{Mat93} and \citet{Mag01b} simulated the three-dimensional
magnetohydrodynamics (MHD) of the emerging magnetic flux.
\citet{Noz05} and \citet{Mur06} also performed the MHD
simulations of flux emergence with the sheared or twisted flux tube.
They found that the flux tube with shear or twist emerges faster than that
without any shear and twist.

\citet{Noz92} performed the MHD simulation of flux
emergence in a sheet geometry.
The initial stable flux sheet in the convective zone was perturbed with various
wavelengths, which correspond to the convective motion.
They found that irrespective of the wavelength of initial perturbation, a finite
``most unstable wavelength'' is excited.
This wavelength (2-4 Mm) is inherent in the Parker instability \citep{Par66}.
As a result, the flux sheet is undulated and the apexes of the convex
field line (\textsf{$\Omega$}-loops) appear consecutively on the photosphere.
Some of dipped field lines (\textsf{U}-loops) also emerge to form the
regions called ``bald patches \citep{Tit93}''.

On the other hand, observational study on a large EFR and bald patches was
performed by \citet{Par04}. They observed a fairly large ($\sim30$ Mm) EFR with the
magnetogram and found that the emerged field lines undulate vertically.
They revealed that there are many bald patches between the main spots.
These results confirmed that the emerging flux tube does
not rise altogether at a time, but each \textsf{$\Omega$}-loop
component rises individually.
They proposed this model as ``resistive emergence model''.
The distance between two consecutive bald patches is in the range of 2-6 Mm,
which is consistent with a theoretical argument on the flattening of
emerging magnetic field just below the surface and its critical emergence length, first
presented by \citet{Mag01a}.

Recently \citet{Iso07} further developed the simulation performed by
\citet{Noz92} and obtained the result in which the undulated field line caused reconnections with
neighboring \textsf{$\Omega$}-loops and created larger loops.
\citet{Arc09} performed the three-dimensional MHD simulation of the emergence of undulating fieldlines.
These reconnection events are interpreted as a sources of Ellerman bombs
\citep{Ell17, Kur82, Kit83, Mat08a, Mat08b, Wat08}.

The resistive emergence model is applicable to the large EFR.
Furthermore, recently \citet{Ots07} found bald patches inside a small-scale EAR ($\sim 5$
Mm) using Solar Optical Telescope (SOT; \cite{Ich04, Tsu07, Sue07,
Shi07}) aboard \textit{Hinode} \citep{Kos07}.
However, \citet{Cen07} and \citet{Gug08} showed small EARs (2 Mm and 6 Mm,
respectively) without undulated magnetic field.

As stated above, the criteria of the bald patches formation are still uncertain.
Furthermore, in the latest simulation, the footpoints of emerged flux loops
showed converging motion toward the bald patches on the photosphere \citep{Che10}.
Although this converging motion was observed in preceding studies \citep{Str96,
Str99, Ber02, Che08}, statistical analysis on that phenomenon with respect to the
size and magnetic characteristics of the flux emergences has not been done yet.

To clarify the criteria of forming bald patch and converging flow, we performed a
statistical study about the nature of magnetic flux emergences using
SOT. The flux emergence phenomena from small EARs to large EFRs observed by SOT were
investigated on their morphological and magnetical characteristics.
Furthermore, we derived the the relations between the total magnetic flux and
the maximum spatial size, the flux growth rate and the mean separation speed 
of the emergence event to clarify how the total flux amount controls the entire
evolution of the emergence.

\section{Observation and Data Reduction}
\label{seq:2}
\subsection{Observation and Data Selection}
\textit{Hinode} satellite has observed solar surface for over 4 years with SOT. 
SOT has Broadband Filter Imager (BFI) and Narrowband Filter Imager (NFI).
Ca\emissiontype{II} H (3968.5 {\AA}) filtergrams were taken by BFI with the bandpass of 3 {\AA}.
Fe\emissiontype{I} (6302 {\AA}) and Na\emissiontype{I} D (5896 {\AA}) 
polarimetric data in solar photosphere and chromosphere were observed with NFI.
To search for the emerging flux phenomena, we used \textit{Hinode} daily quicklook movies
\footnote{http://hinode.nao.ac.jp/QLmovies/}.
Among the possible candidates, we selected 101 emerging flux
phenomena according to the criteria as follows:
(1) conspicuous presence of separating bright points in Ca\emissiontype{II} H
image and/or opposite polarities in Fe\emissiontype{I} or Na\emissiontype{I} D image,
(2) location fairly inside the solar limb
($\alpha=\arcsin(r/R)\le\timeform{70D}$, where $r$ is the distance from solar disk center to the location of the event
and $R$ is the solar radius)
(3) observational time span longer than 1 hour, and
(4) observational cadence higher than 10 minutes.
The period in which we studied is between 2006 November 26 to 2010 August 23.
The detailed observational information of the emerging flux is available in the
associated electronic tables\footnote{Before the publication of this paper, the
table is shown in http://www.kwasan.kyoto-u.ac.jp/$\sim$otsuji/electronic\_table.pdf.
After publication the table will be uploaded in PASJ site.}.

\subsection{Data Reduction}
In this section we give the description of the data reduction with the
data of EFR 20061201 (\# 2 in the electronic table) as an example.
First, dark-current subtraction and flat fielding were
performed on the obtained SOT data in the standard manner.
Then we processed every observed data as described below.

\begin{figure}
\begin{center}
\FigureFile(80mm,80mm){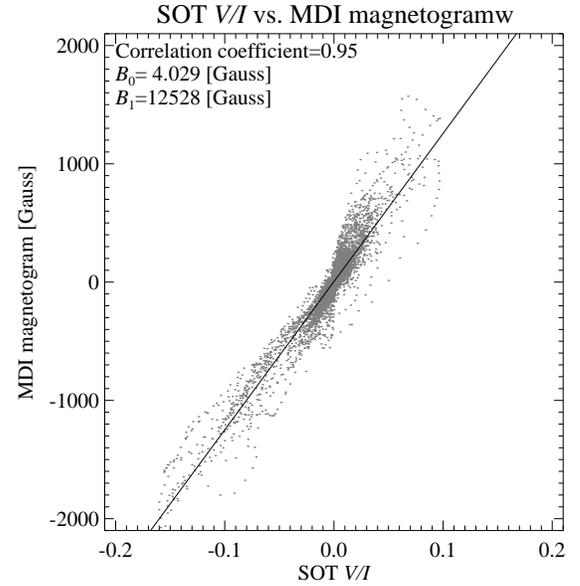}
\end{center}
\caption{Scatter plot of the polarimetric signal $V/I$ of SOT and the field
strength of MDI magnetogram.}
\label{fig:02}
\end{figure}

First, we used \textit{SOHO}-Michelson Doppler Imager (MDI; \cite{Sch95})
magnetogram data to calibrate SOT polarimetric data.
We compensated the differential rotation of two consecutive MDI data observed
before and after the SOT observation and interpolated them by time to estimate
the distribution of magnetic field at the time of SOT observation.
Next we deteriorated SOT polarimetric data with the spatial resolution of MDI
magnetogram ($\timeform{2''}$). Using the deteriorated SOT polarimetric data and MDI magnetogram,
we made a scatter plot of SOT polarimetric signal (Stokes $V/I$) to MDI field
strength (Figure \ref{fig:02}). The correlation coefficient of SOT $V/I$ to MDI field
strength was 0.95. We performed linear fitting on the scatter plot and obtained
the conversion equation from SOT $V/I$ to the photospheric field strength
$B_\mathrm{p}$, expressed as
\begin{equation}
B_\mathrm{p}=B_1\times V/I+B_0.
\end{equation}
In the case of Figure \ref{fig:02}, the offset value $B_0$ and the scaling
factor $B_1$ were 4.029 Gauss and 12528 Gauss, respectively.
Note that the scatter of data points in the figure is mainly due to the Doppler
effect arising from the satellite's orbital motion.
With this method we converted the SOT polarimetric signals to the magnetic flux
densities for all the samples.

\begin{figure}
\begin{center}
\FigureFile(80mm,45mm){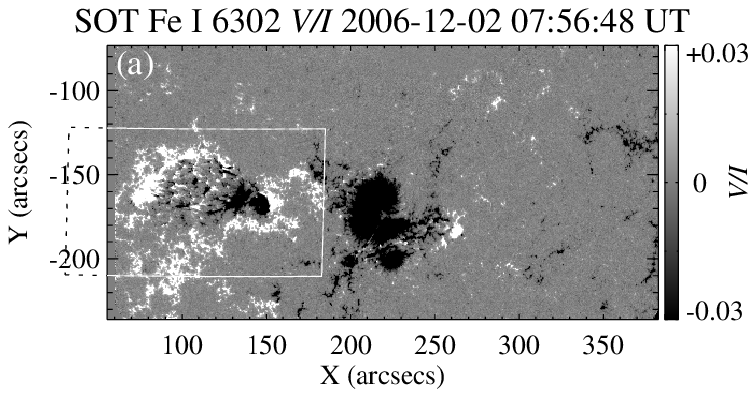}
\FigureFile(80mm,45mm){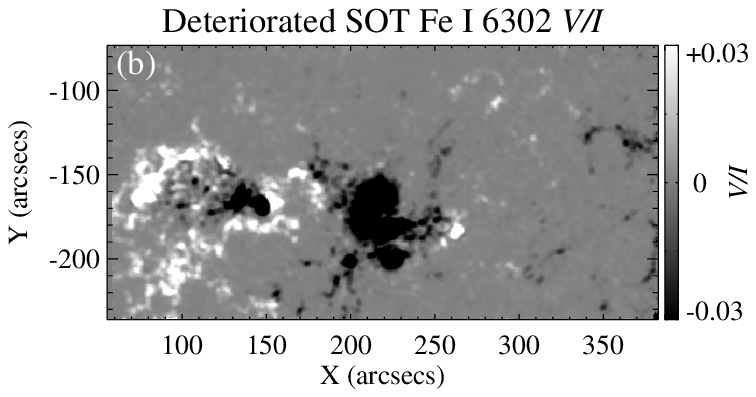}
\FigureFile(80mm,45mm){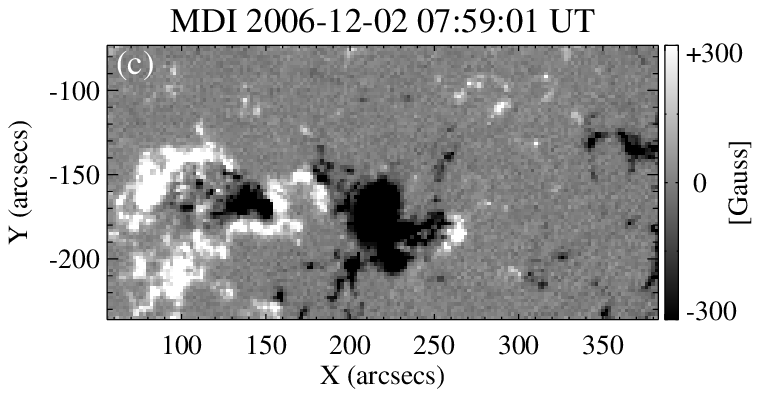}
\end{center}
\caption{(a) Pointing-corrected SOT polarimetric image with the white box
indicating the area of the ``top-view image'' in Figure \ref{fig:04}.
(b) Deteriorated SOT polarimetric image.
(c) Reference \textit{SOHO} MDI image estimated by linear interpolation.}
\label{fig:01}
\end{figure}

In measuring the actual size of the solar features, we compensated
the projection effect.
Then we applied the subsonic
filter of 3 minutes to the SOT Ca\emissiontype{II} H image sequence in order to
suppress chromospheric oscillatory motions (Figure \ref{fig:03}b).

\section{Analysis}
As is stated in the introduction, mutually approaching anti-polarity patches
were observed in undulating resistive emerging phenomena \citep{Str99}.
We studied the formation process of the converging motion between the 
opposite polarities in EFR and EAR by analysing the
morphological, dynamical and magnetic evolutions of our SOT samples.
In this section we introduce our analysis methods using the data of EFR
20061201.

\subsection{Morphological Evolution}
The morphological evolution of the magnetic flux emergence was analysed with two
methods.
One is the method of tracking magnetic elements using local
correlation tracking (LCT), and the other using time-sliced diagram.

\subsubsection{Local Correlation Tracking}
The local correlation tracking (LCT) method is commonly
used to derive the horizontal velocity field \citep{Nov88, Ber98, Mat10}.
For LCT, we used \texttt{flowmap.pro} in SSW
of IDL.
\texttt{Flowmap.pro} calculates the two dimensional vector flowfield by
following the subtiles in the time series of two dimensional images.

\begin{figure*}
\begin{center}
\FigureFile(80mm,50mm){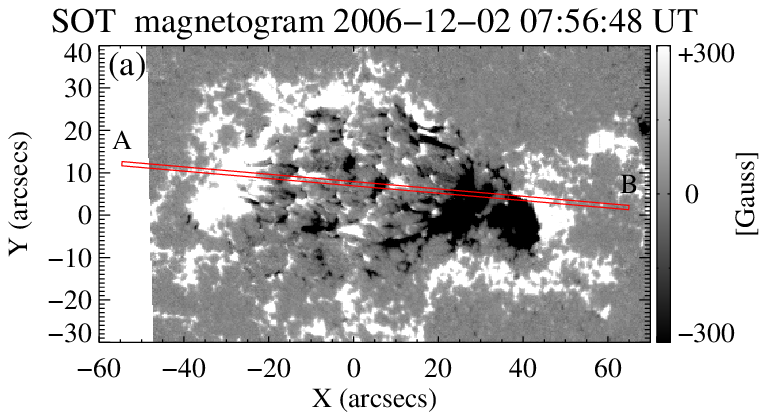}
\FigureFile(80mm,50mm){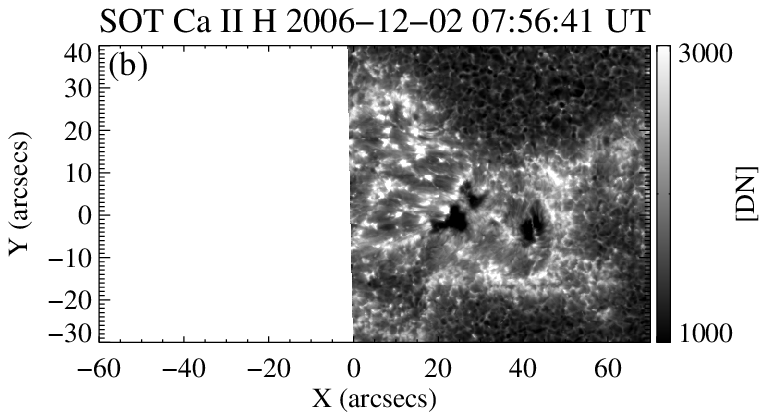}
\FigureFile(80mm,50mm){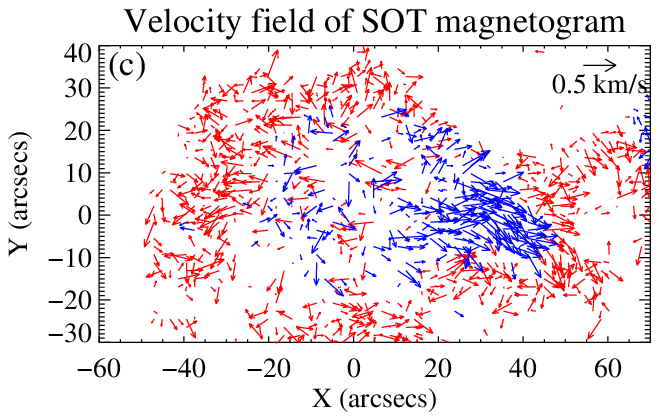}
\FigureFile(80mm,50mm){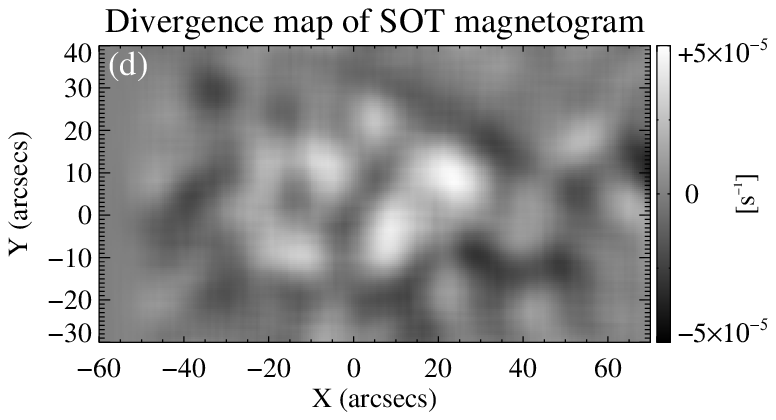}
\FigureFile(80mm,50mm){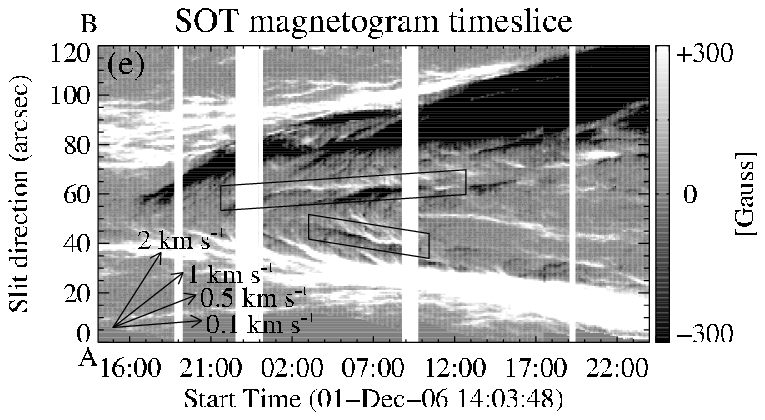}
\FigureFile(80mm,50mm){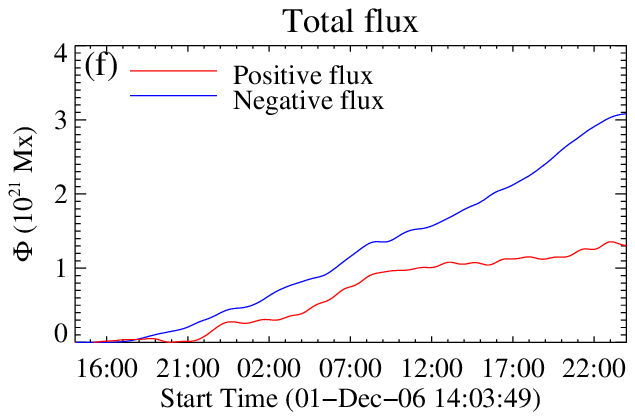}
\end{center}
\caption{EFR 20061201.
(a) SOT Fe\emissiontype{I} magnetogram.
(b) SOT Ca\emissiontype{II} filtergram.
(c) Horizontal velocity field of magnetogram derived by LCT, averaged
with \timeform{1''} and 1 minutes.
Red and blue arrows indicate the velocities of positive and negative
magnetic components.
The threshold field strength of drawing arrows is $\pm50$ Gauss.
(d) The divergence map derived from the horizontal velocity field.
Spatially averaging width is \timeform{10''}.
Temporally averaging period is 10 minutes.
(e) Time-sliced diagram of SOT magnetogram.
The both edge of the slit (A and B) in panel (a) correspond to
\timeform{0''} and \timeform{120''} in the plot, respectively.
The vertical white gaps represent the observation breaks.
The mutually approaching anti-polarities are indicated by black boxes.
(f) The evolution of magnetic flux of the EFR.}
\label{fig:03}
\end{figure*}

To examine the motion of the footpoints of flux tubes for all over the emerging
site, we performed LCT on SOT magnetogram data and obtained velocity field
of moving magnetic elements (Figure \ref{fig:03}c).
The size of tracking subtile for LCT was \timeform{0.5''}.
To reduce the velocity noise due to LCT error, the velocity fields were averaged
both spatially and temporally over the zones of
$\timeform{1''}\times\timeform{1''}$ and 10 minutes, respectively.
The standard deviation of the velocity field inside the data cube of
$\timeform{1''}\times\timeform{1''}\times 10$ minutes was $\sim0.1$ km
s$^{-1}$.

Then we derived divergence of horizontal velocity field (Figure
\ref{fig:03}d).
To emphasize global and sustained flows,
the divergence maps were averaged both spatially and temporally.
The spatial average was performed with the width of \timeform{10''} for large
EFR and \timeform{3''} for small EAR.
The threshold between large EFRs and small EARs was fixed to be \timeform{40''},
which is the upper limit size of ephemeral active regions indicated by
\citet{Har73}. For EFR 20061201, the
spatial averaging box size was \timeform{10''}. The temporally averaging period was taken as 10 minute for all the
events.

\subsubsection{Time-sliced Diagram}
To clarify the dynamics of footpoints more quantitatively, we made
time-sliced diagram of SOT magnetograms (Figure
\ref{fig:03}e).
The spatial slit was located parallel to the axis of the
EFR (shown in Figure \ref{fig:03}a).
From the time-sliced diagram, we derived the maximum distance
$d_\mathrm{max}$ between the main spots, the mean separating speed
$\langle v\rangle$ of the main spots.
Detailed footpoints motions, such as mutually approaching
anti-polarities, were examined in this diagram.

\subsection{Temporal Evolution of Magnetic Field}
To investigate the temporal evolution of emerging magnetic flux, we
measured the total flux within the emerging site.
Positive and negative fluxes were summed up separately.
The total flux of the emerging region $\Phi$ was derived by subtracting the
fluxes at the initial time.
We plotted the variation of positive and negative fluxes, respectively (Figure
\ref{fig:03}f).
From this plot, we derived the maximum amount of unsigned total magnetic flux
$\Phi_\mathrm{max}$ and the unsigned flux growth rate
$\langle\mathrm{d}\Phi/\mathrm{d}t\rangle=\Phi_\mathrm{max}/T$.
The growth rate is defined as the total magnetic flux
$\Phi_\mathrm{max}$ divided by the continuously emerging period $T$.
Note that the unbalance between positive and negative fluxes in the sample plot
was due to the flow of the positive flux out of the field of view.

\section{Results}
First, we introduce the sample results for large EFR and small
EAR at section \ref{sec:4.1} and \ref{sec:4.2}, respectively.
Then the statistical results are shown in section \ref{sec:4.3}.
\subsection{Large EFR 20061201}
\label{sec:4.1}
In Figure \ref{fig:03}a,
there are two main spots aligned east-west direction.
The size of the main spots was \timeform{10''}-\timeform{20''}.
The field strength of the spots was $\pm1.8\times10^3$ Gauss at the maximum.
Although the following spot is missing in the Ca\emissiontype{II} H
image because of the field of view limitation (Figure \ref{fig:03}b),
there are sunspots locating at the same position with the preceding negative
spots in the magnetogram. The velocity field shows prominent outward motions of
the main spots (Figure \ref{fig:03}c).
There are positive divergence areas inside the main spots (Figure
\ref{fig:03}d), which indicates the flux emergences.
From the time-sliced diagram (Figure \ref{fig:03}e), we can estimate the speed
of main spots to be $\sim0.3$ km s$^{-1}$ for each, thus the mean separating
speed $\langle v\rangle$ is about 0.6 km s$^{-1}$.
At the end of the observation period, the distance between two main spots
increased to around \timeform{100''}. We considered this value as the maximum
distance $d_\mathrm{max}$.
The emergence started at 16:00 UT on 1 December and lasted until the observation
end at 24:00 UT on 2 December.
Thus the active emergence period was taken as 32 hours.
The maximum amount of total magnetic flux were
$1.3\times10^{21}$ Mx for positive polarity and $3.1\times10^{21}$ Mx for negative polarity (Figure \ref{fig:03}f).
As the following positive spot flowed out of the field of view, measured
positive flux was less than that of negative one.
Thus we took the maximum negative flux as $\Phi_\mathrm{max}$.
Mean flux growth rate $\langle\mathrm{d}\Phi/\mathrm{d}t\rangle$ for this
event was $9.7\times10^{19}$ Mx hr$^{-1}$.

Let us look at the central part of the region where there are many small
positive or negative magnetic patches (Figure \ref{fig:03}a).
These patches correspond to the Ca\emissiontype{II} H bright points (Figure
\ref{fig:03}b).
Although the magnetic patches seem to move with apparently
random velocities in Figure \ref{fig:03}c, these patches are located in the
converging region of divergence map (Figure \ref{fig:04}d). Thus these magnetic patches accumulated
and stagnated to the localized area.
In fact, the time-sliced diagram shows that these magnetic patches actually
approach to each other with speed of $\sim1$ km s$^{-1}$.
So we identify these mutually approaching area as a \textsf{U}-loop formation in
the EFR.
In the following, we denote the area as the ``stagnation zone (SZ)'',
where the anti-polarities mutually approach and accumulate.

\subsection{Small EAR 20070213}
\label{sec:4.2}
\begin{figure*}
\begin{center}
\FigureFile(80mm,45mm){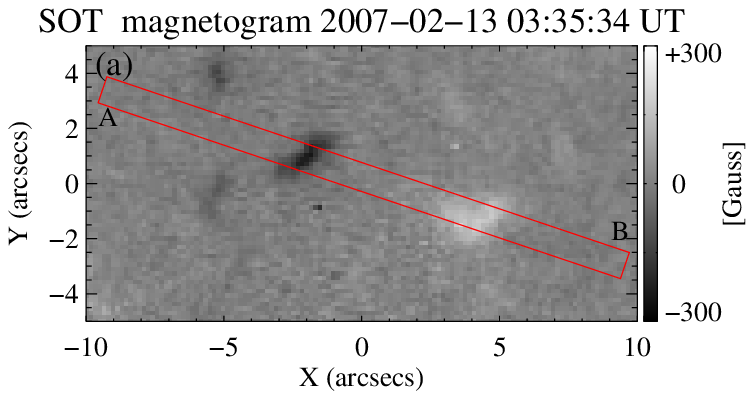}
\FigureFile(80mm,45mm){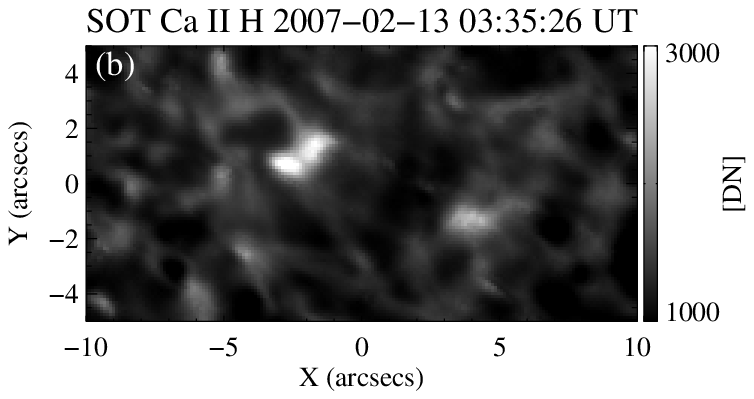}
\FigureFile(80mm,45mm){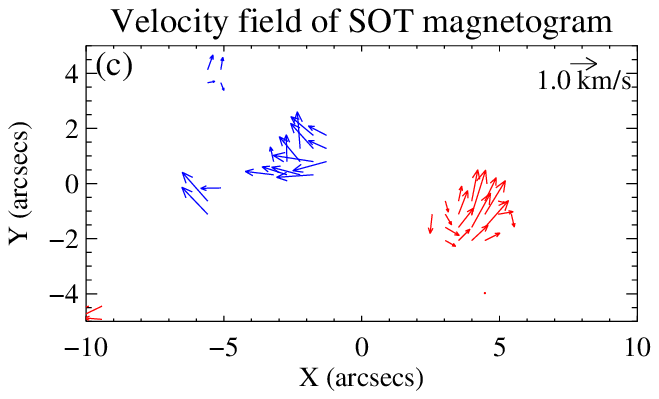}
\FigureFile(80mm,45mm){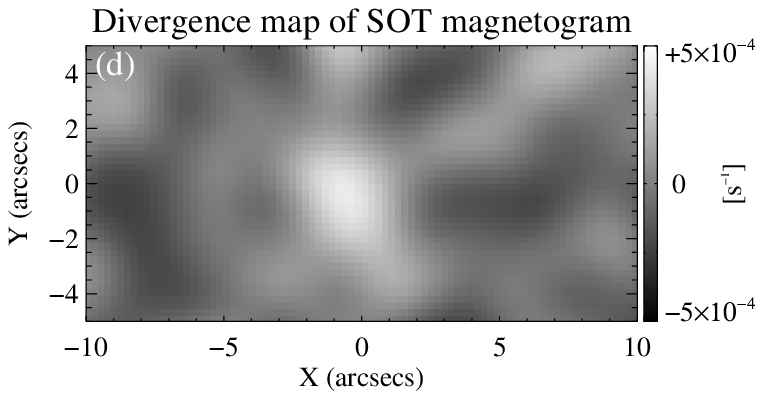}
\FigureFile(80mm,45mm){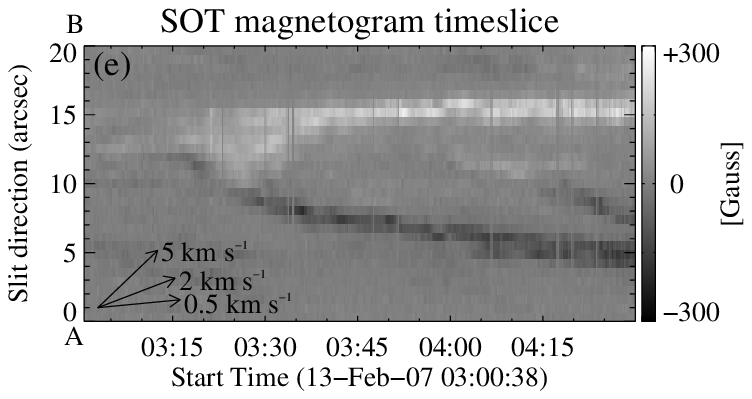}
\FigureFile(80mm,45mm){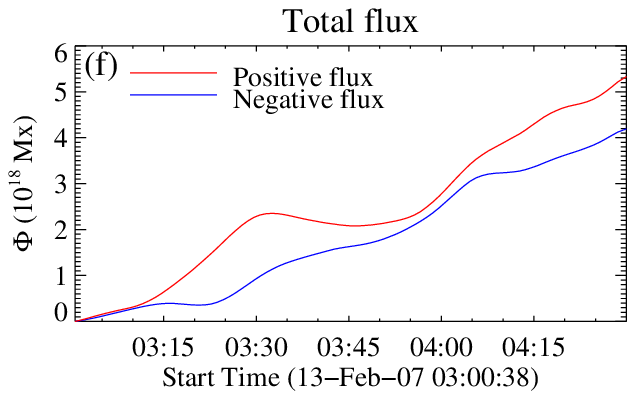}
\end{center}
\caption{EAR 20070213.
(a) SOT Fe\emissiontype{I} magnetogram.
(b) SOT Ca\emissiontype{II} filtergram.
(c) Horizontal velocity field of magnetogram derived by LCT, averaged with
\timeform{1''} and 1 minutes.
Red and blue arrows indicate the velocities of positive and negative
magnetic components.
The threshold field strength of drawing arrows is $\pm50$ Gauss.
(d) The divergence map derived from the horizontal velocity field.
Spatially averaging width is \timeform{3''}.
(e) Time-sliced diagram of SOT magnetogram.
The both edge of the slit (A and B) in panel (a) correspond to
\timeform{0''} and \timeform{20''} in the plot, respectively.
(f) The evolution of magnetic flux of the EAR.}
\label{fig:04}
\end{figure*}

We selected a small-scale magnetic emergence event which emerged on 13
February 2007 (\# 9 in the electronic table) as the sample case of EAR.
Figure \ref{fig:04}a shows the magnetogram of the region. Two magnetic
concentrations in the magnetogram correspond to the Ca\emissiontype{II} H
bright points in Figure \ref{fig:04}b.
The field strength of the two magnetic concentrations was $\pm300$ Gauss at the
maximum.
The velocity field of the small EAR shows separative and anti-clockwise
rotational motion of two magnetic concentrations (Figure \ref{fig:04}c).
Figure \ref{fig:04}d is the divergence map derived from the velocity
field and averaged spatially with \timeform{3''}.
The divergence map shows the positive area at the central region of the EAR,
which indicates that there is no converging flow inside the emerging site.
In the time-sliced diagram we can see the simple separating motion of two main
magnetic concentrations and no stagnation zone (Figure \ref{fig:04}d).
The maximum distance between the main concentration $d_\mathrm{max}$ was
\timeform{10''}. The mean speed of separating motion for main concentrations $\langle
v\rangle$ is about 1.5 km s$^{-1}$.
The temporal evolution of total flux shows that the maximum amounts of total
fluxes are $5.4\times10^{18}$ Mx for positive polarity and $4.2\times10^{18}$ Mx
for negative polarity (Figure \ref{fig:04}f).
We took the maximum positive flux as $\Phi_\mathrm{max}$ for
this event. The time-sliced diagram and the total fluxes evolution plot indicate
that the flux emergence began at 03:15 UT and last until the observation end at 04:30 UT.
Thus the active emergence period was 75 minutes and the mean flux growth
rate $\langle\mathrm{d}\Phi/\mathrm{d}t\rangle$ was $4.3\times10^{18}$ Mx
hr$^{-1}$.

\subsection{Statistical Result}
\label{sec:4.3}
\begin{figure*}
\begin{center}
\FigureFile(80mm,50mm){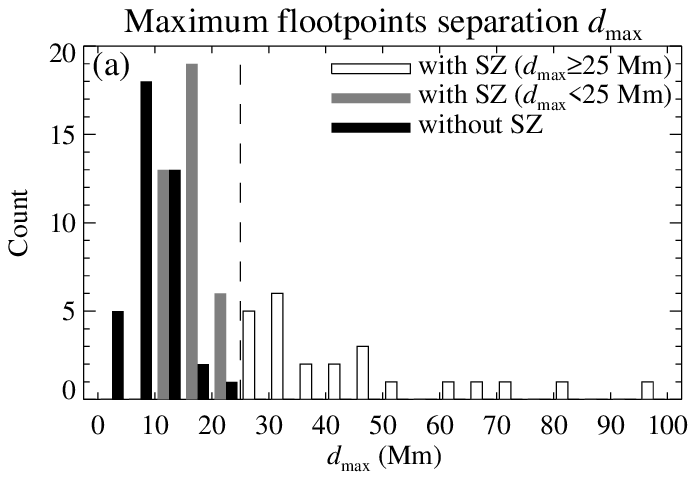}
\FigureFile(80mm,50mm){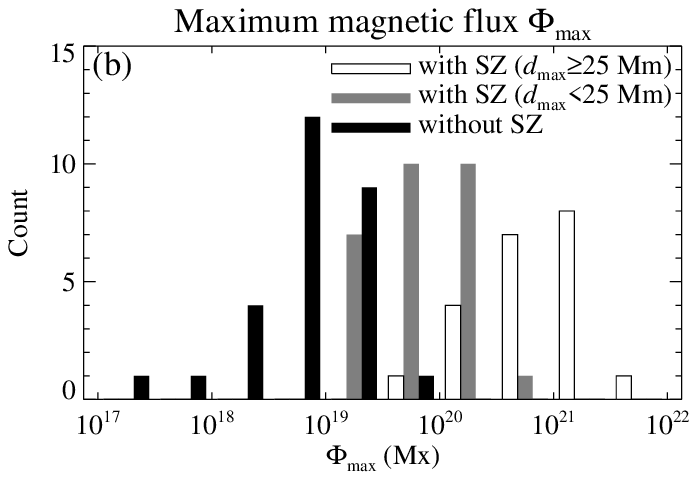}
\FigureFile(80mm,50mm){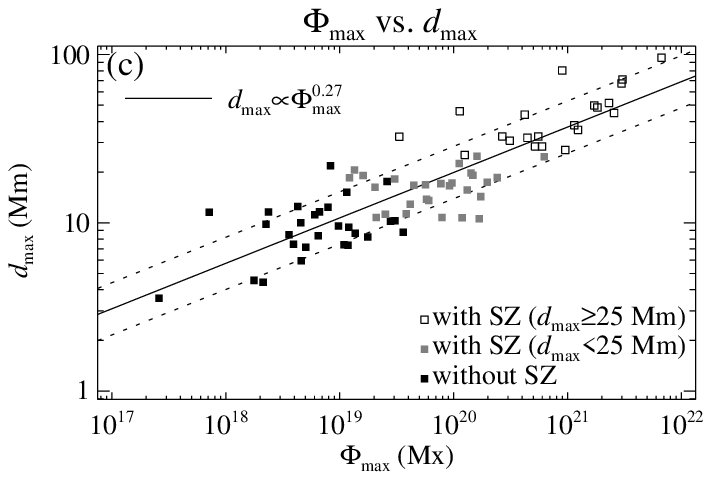}
\FigureFile(80mm,50mm){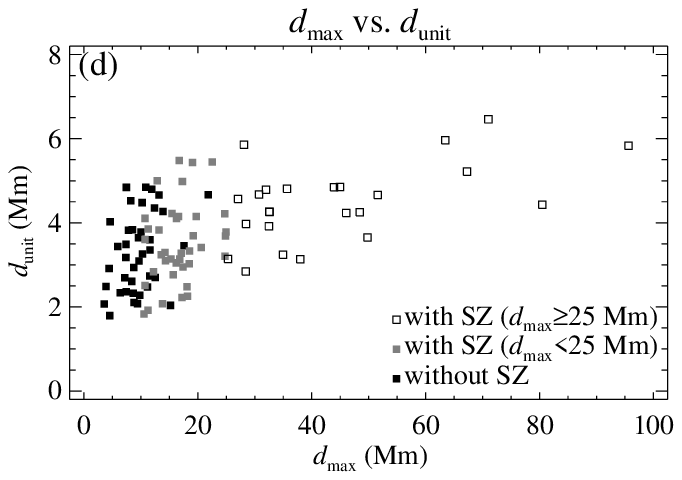}
\FigureFile(80mm,50mm){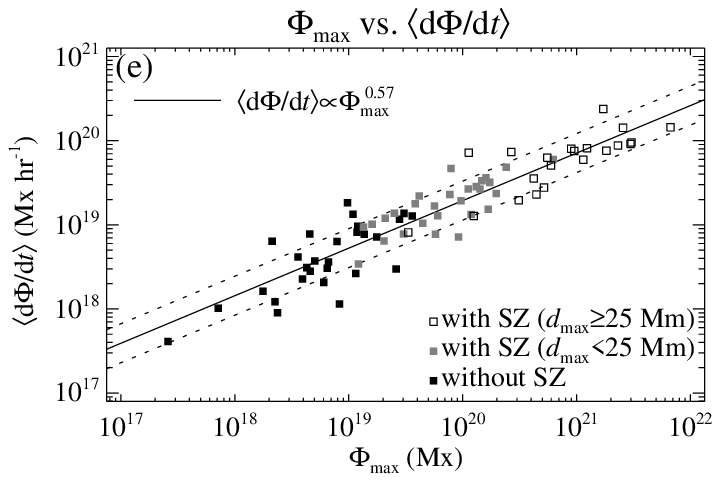}
\FigureFile(80mm,50mm){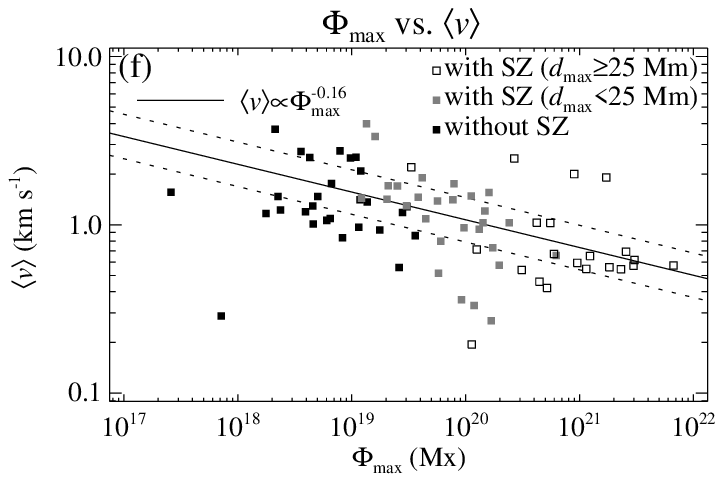}
\end{center}
\caption{Statistical characteristics of magnetic flux emergences.
Data points are plotted with different symbols according to the
association with/without the stagnation zone (SZ).
(a) Histogram of $d_\mathrm{max}$.
(b) Histogram of $\Phi_\mathrm{max}$.
(c) Scatter plot of $\Phi_\mathrm{max}$ and $d_\mathrm{max}$.
The solid line in the plot indicates the distribution relation of
$d_\mathrm{max}=7.9\times10^{-5}\Phi_\mathrm{max}^{0.27}$.
(d) Scatter
plot of $d_\mathrm{max}$ and $d_\mathrm{unit}$.
(e) Scatter plot of $\Phi_\mathrm{max}$ and 
$\langle\mathrm{d}\Phi/\mathrm{d}t\rangle$.
The solid line in the plot indicates the distribution relation of
$\langle\mathrm{d}\Phi/\mathrm{d}t\rangle=9.6\times10^{7}\Phi_\mathrm{max}^{0.57}$.
(f) Scatter plot of $\Phi_\mathrm{max}$ and $\langle v\rangle$.
The solid line in the plot indicates the distribution relation of
$\langle v\rangle=2.1\times10^{3}\Phi_\mathrm{max}^{-0.16}$.
The dashed lines in the panels represent the deviation (1$\sigma$) of the
fitting plots.}
\label{fig:05}
\end{figure*}

Table \ref{tab:01} in Appendix \ref{seq:a1} shows the measured quantities of all
the samples. If there was no magnetic observation, Ca\emissiontype{II} H data were used
to derive $d_\mathrm{max}$ and $\langle v\rangle$, and to judge the
existence of SZs.
There is no data of the total flux $\Phi_\mathrm{max}$ and the flux growth rate
$\langle\mathrm{d}\Phi/\mathrm{d}t\rangle$ for the observations without the
magnetogram.
Figure \ref{fig:05} shows the statistical characteristics of
measured quantities.

\subsubsection{Existence of SZ}
In Figure \ref{fig:05}a, the histogram of the maximum separation distance
$d_\mathrm{max}$ indicates that SZs are rarely found in small $d_\mathrm{max}$
regions.
The flux emergence phenomena with the separation size of more than 25 Mm always
have SZs. Below the threshold of 25 Mm, there are both EFRs and EARs with/without SZs.
To clarify the conditions to have SZs, we categorized all the events to three
groups as follow.
Group I: maximum separation distance $d_\mathrm{max}\ge 25$ Mm and with SZs.
Group II: maximum separation distance $d_\mathrm{max}<25$ Mm and with SZs.
Group III: maximum separation distance $d_\mathrm{max}<25$ Mm and without SZ.
Group I, II and III are indicated in the Figure \ref{fig:05} by white, gray and
black bars/squares, respectively.
Figure \ref{fig:05}b shows the histogram of the
maximum fluxes $\Phi_\mathrm{max}$, where we present the maximum flux
separately for these three groups.
Group I has the maximum flux of
$\Phi_\mathrm{max}\sim10^{21}$ Mx, while group II and III have
$\Phi_\mathrm{max}\sim10^{20}$ Mx and $\sim10^{19}$ Mx, respectively.
Figure \ref{fig:05}c shows the relation between the maximum flux 
$\Phi_\mathrm{max}$ and the maximum separation distance $d_\mathrm{max}$.
In the plot the three group I, II and III are clearly separated.
The maximum separation distance $d_\mathrm{max}$ depends on the maximum flux
amount $\Phi_\mathrm{max}$.
The scatter plot implies a power-law relation of
\begin{equation}
d_\mathrm{max}=7.9\times10^{-5}\Phi_\mathrm{max}^{0.27},
\label{eq:02}
\end{equation}
where $d_\mathrm{max}$ is in Mm and $\Phi_\mathrm{max}$ is in
Mx.
We can see the trend in which the small EARs have low values of maximum flux
while large EFRs have high values of maximum flux.
The power-law relation is consistent with the result of \citet{Hag01}, although
the index of power was 0.18 instead of 0.27. From the result of categorization, we found that the SZ features are
associated with the magnetic flux emergence of more than around $2\times10^{19}$ Mx.

\subsubsection{Size of Elementary Flux Emergence}
\label{sec:4.3.2}
We also measured the typical size of elementary structures of emergence
$d_\mathrm{unit}$ defined as the distance between two footpoints of
individual \textsf{$\Omega$}-loops at their emergence period.
Some small footpoints of emerged loops might be finally transported to the
border of supergranules by the local convection.
Others cause cancellation between the opposite polarities and disappear,
which enlarges the distance between two footpoints of
individual \textsf{$\Omega$}-loops, (i.e. $d_\mathrm{unit}$).
Thus the $d_\mathrm{unit}$ varies with time.
For the accurate descriptions, we adopted the $d_\mathrm{unit}$ at the epoch
when the \textsf{$\Omega$}-loops were observed as Ca\emissiontype{II} H filaments
($\sim$10 minutes after the start of the emergence; \cite{Ots07}).
Figure \ref{fig:05}d shows the scatter plot of $d_\mathrm{max}$ and $d_\mathrm{unit}$,
which suggest that $d_\mathrm{unit}$ takes the values in the range of 2-6 Mm irrespective of $d_\mathrm{max}$.
Thus elementary and unitary \textsf{$\Omega$}-loops in any emerging flux region has a common size of around 4 Mm,
which is consistent with the most unstable wavelength (2-4 Mm) of Parker instability and preceding
observation/simulation studies \citep{Mag01a, Par04, Iso07}.

\subsubsection{Magnetic Flux Evolution}
The relation between the maximum flux $\Phi_\mathrm{max}$ and the flux growth
rate $\langle\mathrm{d}\Phi/\mathrm{d}t\rangle$ is shown in Figure \ref{fig:05}e.
In the scatter plot, the data points distribute along the relation of
\begin{equation}
\left\langle\frac{\mathrm{d}\Phi}{\mathrm{d}t}\right\rangle=\frac{\Phi_\mathrm{max}}{T}=9.6\times10^{7}\Phi_\mathrm{max}^{0.57},
\label{eq:03}
\end{equation}
where $T$ is the emergence duration, $\langle\mathrm{d}\Phi/\mathrm{d}t\rangle$
is in Mx hr$^{-1}$ and $\Phi_\mathrm{max}$ in Mx.
From equation (\ref{eq:03}), we can derive the emergence
duration $T$ in the unit of hour as a function of $\Phi_\mathrm{max}$, which
is
\begin{equation}
T=1.03\times10^{-8}\Phi_\mathrm{max}^{0.43}.
\label{eq:04}
\end{equation}
Equation (\ref{eq:04}) indicates that an emergence event with large maximum flux
shows relatively rapid magnetic flux growth.
According to equation (\ref{eq:04}), when a flux tube of $\Phi_0$ emerges with
$T_0$, the flux tube with $2\Phi_0$ emerges with $T_1=2^{0.43}T_0\approx1.4T_0$.
$T$ does not depend linearly but non-linearly on $\Phi_\mathrm{max}$.
A tube with more magnetic flux emerges with less time than in the case of linear
dependency.
While the equation (\ref{eq:04}) is consistent with previous
observations such as \citet{Zwa87} and \citet{Hag01}, the empirical relation
(\ref{eq:04}) was first derived with wide range of magnetic parameters by
\textit{Hinode} high-resolution samples.

\subsubsection{Relation between Footpoints Separating Speed and Maximum Flux}
Figure \ref{fig:05}f presents the relation between
$\Phi_\mathrm{max}$ and $\langle v\rangle$, which indicates that the 
larger size EFRs show the separating speed less than 1 km s$^{-1}$, while the
small scale EARs footpoints separate with various speed up to 4 km
s$^{-1}$.
The mean separating speed $\langle v\rangle$ can be written as
\begin{equation}
\langle v\rangle=\frac{d_\mathrm{max}}{T}.
\label{eq:05}
\end{equation}
From equations (\ref{eq:02}) and (\ref{eq:04}), equation (\ref{eq:05}) reduces
to
\begin{equation}
\langle v\rangle=2.1\times10^{3}\Phi_\mathrm{max}^{-0.16},
\label{eq:06}
\end{equation}
where $\langle v\rangle$ is in km s$^{-1}$ and $\Phi_\mathrm{max}$ in Mx.
Equation (\ref{eq:06}) indicates that the footpoints of emerged flux tube with
less magnetic flux separate each other with larger speed.
This relation is plotted with solid line in Figure \ref{fig:05}e, which is
consistent with the observed values.

\section{Discussion}
\subsection{Size and Flux Dependence of SZ Formation}
The SZ features are associated with the magnetic flux emergence of more
than around $2\times10^{19}$ Mx.
\citet{Mag03} performed three-dimensional MHD simulation of emerging
magnetic flux and suggested that emerging field lines take the
evolutionary path of a simple expansion if they emerge with
a large aspect ratio (the ratio of their height to their footpoint
distance); otherwise, field lines are inhibited from
expanding and they show an undulating behavior (i.e. SZ formation).
Our results provide a new criterion of the total magnetic flux
regarding the formation of SZs.

\subsection{Flux Dependence of the Spatial Size of Flux Emergences}
Another notable result is that the flux emergence
phenomena with/without SZ follow the relation between the total flux and the
maximum spatial size as described in equation (\ref{eq:02}).
This relation is derived from wider range of magnetic parameters than
the previous studies \citep{Bru67, Zwa87, Har73, Har75}.
There authors gave the result for only narrow range of magnetic parameters.
We present a comprehensive result on this relation compared to the previous
works.

\begin{figure}
\begin{center}
\FigureFile(80mm,80mm){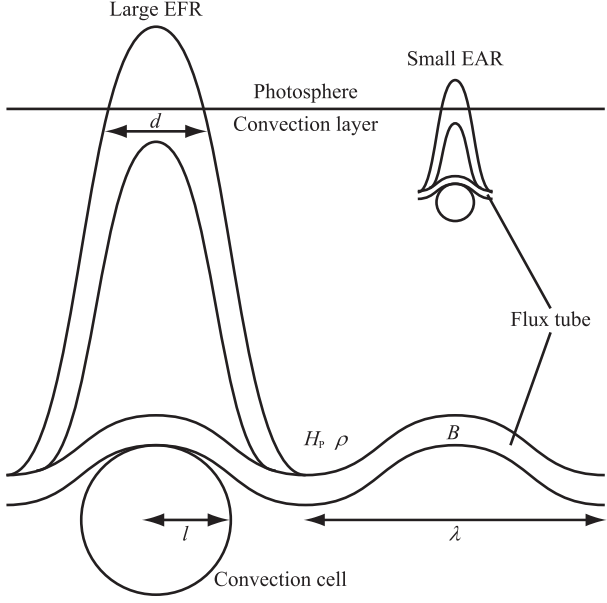}
\end{center}
\caption{
Schematic image of emerging flux tubes.
$d$ is the distance between the main spots.
$l$ is the mixing length.
$H_\mathrm{P}$ is the pressure scale height at the initial
depth.
$B$ is the field strength in the flux tube.
$\rho$ is the mass density around the flux tube.
$\lambda$ is the most unstable wavelength of Parker instability.}
\label{fig:06}
\end{figure}

Let us try to derive the power-law relation between the maximum flux
$\Phi_\mathrm{max}$ and the maximum separation distance $d_\mathrm{max}$,
\begin{equation}
d_\mathrm{max}\propto\Phi_\mathrm{max}^{\alpha_1}
\end{equation}
from the viewpoint of simple dimensional analysis.
Figure \ref{fig:06} shows the schematic image of flux emergence from the
convection layer.
First, $d_\mathrm{max}$ is estimated as follows.
Initially horizontal flux tube in the convection layer will rise with the
typical length $\lambda=10\sim20H_\mathrm{P}$ where $H_\mathrm{P}$ is the local
pressure scale height.
The maximum separation distance $d_\mathrm{max}$ between
two main spots depends on the most unstable wavelength of Parker instability
$\lambda$ at the initial depth of the flux tube,
\begin{equation}
d_\mathrm{max}\propto\lambda\propto H_\mathrm{P}.
\label{eq:08}
\end{equation}
Thus $d_\mathrm{max}$ is proportional to $H_\mathrm{P}$ at the depth where
initially the flux tube locates.

Next the total flux $\Phi_\mathrm{max}$ is estimated as follows.
From equipartition arguments, the magnetic and kinetic energy in the solar
convection layer will balance to each other,
\begin{equation}
\frac{B^2}{8\pi}\simeq\frac{1}{2}\rho v_\mathrm{conv}^2,
\label{eq:09}
\end{equation}
where $B$, $\rho$ and $v_\mathrm{conv}$ are the field strength inside the
flux tube, the mass density around the flux tube and the mean convection velocity.
From mixing length theory \citep{Sti89}, $v$ is given as
\begin{equation}
v_\mathrm{conv}\propto\sqrt{H_\mathrm{p}}.
\label{eq:10}
\end{equation}
From equations (\ref{eq:09}) and (\ref{eq:10}),
\begin{equation}
B\propto\sqrt{\rho H_\mathrm{P}}.
\label{eq:11}
\end{equation}
Now we assume that the solar convection layer can be approximated by an
adiabatically stratified atmosphere \citep{Fou04},
\begin{equation}
T\propto\rho^{\gamma-1}.
\end{equation}
$T$ and $\gamma$ are temperature and adiabatic index
$\gamma=c_\mathrm{P}/c_\mathrm{V}$, where $c_\mathrm{P}$ and $c_\mathrm{V}$ are
the specific heats at constant pressure and volume, respectively.
The local scale
height $H_\mathrm{P}$ is proportional to the temperature $T$, thus
\begin{equation}
\rho\propto T^\frac{1}{\gamma-1}\propto
H_\mathrm{P}^\frac{1}{\gamma-1}.
\label{eq:13}
\end{equation}
From equation (\ref{eq:11}) and (\ref{eq:13}),
\begin{equation}
B\propto H_\mathrm{P}^\frac{\gamma}{2(\gamma-1)}.
\end{equation}
Let us think about the flux tube width $w$.
If the $w$ is much larger or smaller than the local mixing length $l\propto
H_\mathrm{P}$, the flux tube will be disintegrated by the convection flows or accumulated at the
convection boundary.
Thus the flux tube width is expected to be comparable to the mixing length,
\begin{equation}
w\sim l\propto H_\mathrm{P}.
\end{equation}
Thus the total flux $\Phi_\mathrm{max}$ can be estimated as 
\begin{equation}
\Phi_\mathrm{max}\sim w^2B\propto H_\mathrm{P}^\frac{5\gamma-4}{2(\gamma-1)}.
\label{eq:16}
\end{equation}
From equation (\ref{eq:08}) and (\ref{eq:16}),
\begin{equation}
d_\mathrm{max}\propto \Phi_\mathrm{max}^\frac{2(\gamma-1)}{5\gamma-4}.
\end{equation}
Thus the power-law index of the relation between the maximum flux
$\Phi_\mathrm{max}$ and the maximum separation distance $d_\mathrm{max}$ is
derived to be $\alpha_1=\frac{2(\gamma-1)}{5\gamma-4}$.
For example, $\alpha_1$ is 0.30 with the adiabatic index $\gamma=5/3$ (ideal
gas case realised in deep convective layers).
With $\gamma\sim4/3$ at near the solar surface where the ionization status is
changing rapidly \citep{Bha05}, $\alpha_1$ is 0.25.
These calculated values are comparable to the observed value $\alpha_1=0.27$.
From the argument above, we get the view of the emergence depicted as in
Figure \ref{fig:06}. Magnetic tubes of large flux are anchored in deep layers
and appear with large separation between two main spots on the solar photosphere.

\subsection{Flux Growth Rate}
We derived the relation between the maximum flux 
$\Phi_\mathrm{max}$ and the flux growth rate
$\langle\mathrm{d}\Phi/\mathrm{d}t\rangle$ to be equation (\ref{eq:03}).

\begin{figure}
\begin{center}
\FigureFile(80mm,60mm){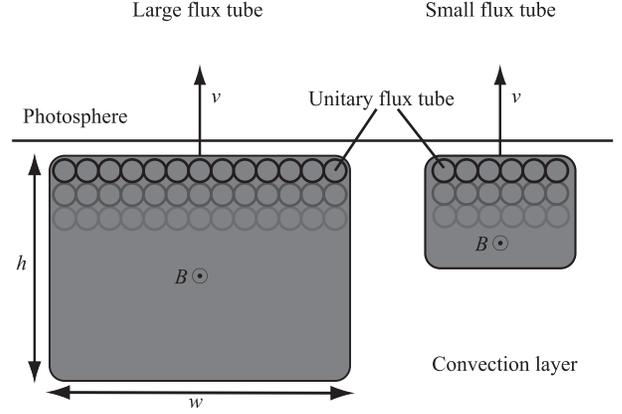}
\end{center}
\caption{Schematic image of flux tube underneath solar photosphere.
The gray areas represent the cross-section of flux tubes with the field strength
of $B$. $w$ and $h$ are horizontal and vertical width of the flux tube, respectively.
$v$ is the rise velocity of the flux tubes at the photosphere.
Small circles in the flux tube represent unitary and elementary flux tube
found in Section \ref{sec:4.3.2}.}
\label{fig:07}
\end{figure}

By a simple model of an emerging tube with uniform magnetic flux density,
let us try to derive the power-law relation between the maximum flux and the
flux growth rate,
\begin{equation}
\langle\mathrm{d}\Phi/\mathrm{d}t\rangle\propto\Phi_\mathrm{max}^{\alpha_2}.
\end{equation}
Figure \ref{fig:07} shows the schematic image of the flux tubes just beneath the
photosphere.
Around the photosphere, the plasma $\beta$ is almost 1.
Thus the magnetic pressure $B^2/8\pi\sim P=\mathrm{const.}$
where $P$ is the gas pressure at the photosphere.
So we can think that flux densities are nearly constant irrespective of the spatial size or total flux of the
magnetic tube,
\begin{equation}
B=\mathrm{const.}
\label{eq:19}
\end{equation}

The rise velocity $v$ of the flux tube is estimated as follows.
When the apex of flux tube reaches to the underneath the solar surface,
the rise motion is suppressed and the tube top becomes flattened \citep{Mag01a}.
Our observation (Section \ref{sec:4.3.2}) showed that the emergence occurs in
unitary form irrespective of total magnetic flux.
In these situations the rise velocity $v$ from the photosphere will not depend
upon the total magnetic flux of the tube,
\begin{equation}
v=\mathrm{const.}
\end{equation}
Assuming that the aspect ratio of flux tube width $h/w$ is constant just
beneath the photosphere,
\begin{equation}
\frac{h}{w}=\mathrm{const.}
\end{equation}
where $w$ and $h$ are horizontal and vertical width of the flux tube,
respectively.

The flux growth rate and the total flux are described as
\begin{equation}
\langle\mathrm{d}\Phi/\mathrm{d}t\rangle=wvB\propto w,
\label{eq:22}
\end{equation}
\begin{equation}
\Phi_\mathrm{max}=whB\propto wh\propto w^2,
\label{eq:23}
\end{equation}
respectively.
From equation (\ref{eq:22}) and (\ref{eq:23}), the relation between
$\Phi_\mathrm{max}$ and $\langle\mathrm{d}\Phi/\mathrm{d}t\rangle$ are written as
\begin{equation}
\langle\mathrm{d}\Phi/\mathrm{d}t\rangle\propto\Phi_\mathrm{max}^\frac{1}{2}.
\end{equation}
Thus the power-law index is
\begin{equation}
\alpha_2=0.5,
\end{equation}
which is consistent with the observed value 0.57.

In this discussion we have not considered the factors such as magnetic
field stratification inside the flux tube, realistic aspect ratio of tube and
so on.
If we include these factors then we can get more realistic interpretation of
equation (\ref{eq:03}).

\section{Summary}
We investigated the morphological, dynamical and magnetical characteristics
of various flux emergence phenomena using high-resolution \textit{Hinode} SOT
data. To estimate the magnetic field density of SOT data we used \textit{SOHO} MDI
magnetogram data for the calibration.
From 101 samples of flux emergence events, we derived the total flux,
flux growth rate, maximum separation and mean separation speed.
The SZ features are associated with the magnetic flux emergence of more
than around $2\times10^{19}$ Mx.
The magnetic flux growth rate, emergence duration and mean separation speed 
were found to follow the power-law functions of the total magnetic flux with 
the indices of 0.57, 0.43 and -0.16, respectively.
The typical size of elementary emergence structures is around 4 Mm, which is
consistent with the most unstable wavelength (2-4 Mm) of Parker instability.
The mean separating speed $\langle v\rangle$ decreases with larger magnetic
flux.

We got a physical view of solar flux emergence that emerging magnetic fields
float and evolve balancing to the surrounding turbulent atmosphere from the discussion on the
observed power-law relations.
These observational results should be verified by future numerical studies.
Possible influence of twisting or pre-existing magnetic field could be studied
with the data of horizontal magnetic field in the emerging site and
will be reported in the near future.

\bigskip
We are grateful to the operation team of \textit{Hinode} and
\textit{SOHO}.
This work was supported by the Grant-in-Aid for the Global COE
Program ``the Next Generation of Physics, Spun from Universality and
Emergence'' from the Ministry of Education, Culture, Sports, Science and
Technology (MEXT) of Japan, \textit{Hinode} is a Japanese mission developed and
launched by ISAS/JAXA, collaborating with NAOJ as a domestic partner, NASA and
STFC (UK) as international partners. Scientific operation of the \textit{Hinode}
mission is conducted by the \textit{Hinode} science team organized at ISAS/JAXA. This
team mainly consists of scientists from institutes in the partner countries.
Support for the post-launch operation is provided by JAXA and NAOJ (Japan),
STFC (U.K.), NASA, ESA, and NSC (Norway).

\appendix
\section{Measured Quantities}
\label{seq:a1}
\begin{table*}
\caption{Measured quantities of 101 samples$^*$}
\label{tab:01}
\begin{center}
\begin{tabular}{rcrrrrc|rcrrrrc}
\hline
\#&
Date&
\multicolumn{1}{c}{$d_\mathrm{max}$}&
\multicolumn{1}{c}{$\langle v\rangle$}&
\multicolumn{1}{c}{$\Phi_\mathrm{max}$}&
\multicolumn{1}{c}{$\langle\mathrm{d}\Phi/\mathrm{d}t\rangle$}&
\multicolumn{1}{c}{SZ}&
\#&
Date&
\multicolumn{1}{c}{$d_\mathrm{max}$}&
\multicolumn{1}{c}{$\langle v\rangle$}&
\multicolumn{1}{c}{$\Phi_\mathrm{max}$}&
\multicolumn{1}{c}{$\langle\mathrm{d}\Phi/\mathrm{d}t\rangle$}&
\multicolumn{1}{c}{SZ}\\
\hline
1&20061126&15.2&0.97&1.2e19&2.6e18&N&51&20080807&7.4&1.87&---&---&N\\
2&20061201&71.0&0.62&3.0e21&9.5e19&Y&52&20080809&3.9&3.09&---&---&N\\
3&20061209&16.6&0.36&9.2e19&7.1e18&Y&53&20081014&12.2&1.07&---&---&Y\\
4&20061226&24.9&0.74&---&---&Y&54&20081018&7.5&1.20&3.9e18&2.3e18&N\\
5&20070104&17.4&0.58&2.0e20&2.4e19&Y&55&20081019&8.4&1.09&6.5e18&3.0e18&N\\
6&20070117&17.3&0.26&---&---&Y&56&20090118&14.3&0.73&1.7e20&3.2e19&Y\\
7&20070204&10.7&0.33&1.2e20&1.3e19&Y&57&20090226&17.6&0.56&2.6e19&3.0e18&N\\
8&20070205&19.2&1.21&1.5e20&3.3e19&Y&58&20090314&10.9&1.45&---&---&N\\
9&20070213&7.2&1.47&5.0e18&3.7e18&N&59&20090402&8.8&0.86&3.6e19&1.3e19&N\\
10&20070219&8.5&2.72&3.6e18&4.1e18&N&60&20090426&18.5&1.44&1.2e19&3.4e18&Y\\
11&20070308&16.8&1.38&5.7e19&1.7e19&Y&61&20090601&32.6&2.48&2.7e20&7.3e19&Y\\
12&20070328&7.8&0.29&---&---&N&62&20090603&30.7&0.54&3.1e20&2.0e19&Y\\
13&20070331&11.6&1.76&6.6e18&3.6e18&N&63&20090623&7.4&2.52&1.1e19&1.3e19&N\\
14&20070414&10.7&1.70&2.1e19&1.2e19&Y&64&20090704&63.4&0.35&---&---&Y\\
15&20070419&16.7&1.09&4.5e19&1.0e19&Y&65&20090707&51.6&0.54&2.3e21&8.8e19&Y\\
16&20070601&15.3&0.50&---&---&Y&66&20090821&5.9&1.01&4.6e18&2.8e18&N\\
17&20070603&14.2&0.73&---&---&Y&67&20090928&10.2&1.18&2.8e19&1.2e19&N\\
18&20070702&8.2&0.93&1.8e19&7.2e18&N&68&20091009&9.8&1.47&2.3e18&1.2e18&N\\
19&20070807&32.0&0.46&4.4e20&2.3e19&Y&69&20091015&9.4&2.09&1.2e19&9.6e18&N\\
20&20070825&12.9&1.90&4.2e19&2.2e19&Y&70&20091019&10.0&1.29&4.5e18&7.8e18&N\\
21&20070918&9.6&2.49&9.7e18&1.8e19&N&71&20091023&34.9&0.44&---&---&Y\\
22&20070929&20.6&3.99&1.3e19&9.4e18&Y&72&20091025&9.7&1.00&---&---&N\\
23&20071001&22.5&1.48&1.1e20&2.7e19&Y&73&20091026&49.8&1.91&1.7e21&2.4e20&Y\\
24&20071008&11.2&1.70&2.5e19&1.4e19&Y&74&20091207&11.2&1.06&6.0e18&2.1e18&N\\
25&20071009&16.3&1.42&2.0e19&6.4e18&Y&75&20091215&80.5&2.00&9.0e20&8.0e19&Y\\
26&20071015&6.4&0.59&---&---&N&76&20091226&19.7&1.03&1.4e20&2.7e19&Y\\
27&20071020&8.8&0.63&---&---&N&77&20091229&32.6&1.03&5.5e20&6.3e19&Y\\
28&20071026&8.7&1.36&1.4e19&7.7e18&N&78&20091230&67.3&0.57&3.0e21&9.1e19&Y\\
29&20071108&4.4&3.69&2.1e18&6.4e18&N&79&20100108&28.4&0.67&6.0e20&5.1e19&Y\\
30&20071111&19.1&3.35&1.6e19&1.0e19&Y&80&20100109&45.0&0.69&2.6e21&1.4e20&Y\\
31&20071117&7.4&1.41&1.2e19&8.1e18&N&81&20100110&95.6&0.57&6.7e21&1.4e20&Y\\
32&20071123&24.8&0.63&---&---&Y&82&20100124&15.7&0.94&1.3e20&2.8e19&Y\\
33&20071127&12.5&2.51&4.3e18&3.1e18&N&83&20100217&25.3&0.71&1.3e20&1.3e19&Y\\
34&20071129&13.2&1.11&---&---&Y&84&20100221&10.7&1.75&7.9e19&4.7e19&Y\\
35&20071207&13.2&0.33&---&---&N&85&20100222&32.5&2.20&3.3e19&8.1e18&Y\\
36&20071208&35.6&0.65&1.2e21&8.1e19&Y&86&20100313&11.6&1.22&2.4e18&9.0e17&N\\
37&20071209&17.2&0.96&9.6e19&1.9e19&Y&87&20100322&4.5&1.17&1.8e18&1.6e18&N\\
38&20071210&24.9&1.56&1.6e20&3.6e19&Y&88&20100325&11.3&1.46&3.8e19&1.8e19&Y\\
39&20071211&27.1&0.59&9.5e20&7.5e19&Y&89&20100414&10.3&1.28&3.1e19&1.4e19&N\\
40&20071212&15.4&0.75&---&---&Y&90&20100612&46.0&0.19&1.1e20&7.2e19&Y\\
41&20071213&24.7&0.66&6.3e20&6.0e19&Y&91&20100613&28.1&0.65&---&---&Y\\
42&20080105&11.9&0.45&---&---&N&92&20100618&13.8&0.51&5.8e19&7.7e18&Y\\
43&20080106&13.9&1.00&---&---&N&93&20100619&38.0&0.55&1.1e21&5.9e19&Y\\
44&20080113&18.1&1.09&---&---&Y&94&20100723&18.5&1.03&2.4e20&4.8e19&Y\\
45&20080124&18.2&1.30&3.0e19&7.8e18&Y&95&20100727&13.6&0.80&6.0e19&1.3e19&Y\\
46&20080419&17.1&1.41&7.7e19&2.3e19&Y&96&20100728&48.4&0.56&1.8e21&7.6e19&Y\\
47&20080610&21.8&0.84&8.3e18&1.1e18&N&97&20100805&28.4&0.42&5.2e20&2.8e19&Y\\
48&20080615&16.2&0.33&---&---&Y&98&20100809&10.6&0.27&1.7e20&1.5e19&Y\\
49&20080620&4.6&2.08&---&---&N&99&20100812&11.6&0.29&7.1e17&1.0e18&N\\
50&20080630&3.6&1.56&2.6e17&4.1e17&N&100&20100814&43.9&1.03&4.2e20&3.5e19&Y\\
&&&&&&&101&20100822&12.4&2.75&7.9e18&6.3e18&N\\
\hline
\multicolumn{14}{l}{
\begin{minipage}{17cm}
\footnotesize{$^*$\quad
Column(1): identification number of the flux emergence.
Column(2): date of the flux emergence.
Column(3): maximum distance between two main spots (Mm).
Column(4): mean separating speed of the main spots (km s$^{-1}$).
Column(5): maximum flux increment (Mx). 
Column(6): flux growth rate (Mx hr$^{-1}$).
Column(7): existence of stagnation zone (Yes/No).
\\
---\quad The missing value (---) is due to the event with only
Ca\emissiontype{II} H observation and without magnetogram.}
\end{minipage}
}
\end{tabular}
\end{center}
\end{table*}

\end{document}